\documentclass[twocolumn,pre,showpacs,superscriptaddress]{revtex4}
\usepackage{epsfig}
\usepackage{amsmath}
\usepackage{amsfonts}

\newcommand{\qF}{\mathcal{F}}
\newcommand{\qT}{\mathcal{T}}
\newcommand{\qe}{\varepsilon}

\newcommand{\qo}{\omega}

\begin{document}

\title{Coupling between static friction force and torque}

\author{S\'{\i}lvio R. Dahmen}
\affiliation{Instituto de F\'{\i}sica da UFRGS, CP 15051,
90501--970 Porto Alegre RS Brazil}
\author{Z\'en\'o Farkas}
\affiliation{Department of Physics, Universit\"at
Duisburg-Essen, D-47048 Duisburg, Germany}
\author{Haye Hinrichsen}
\affiliation{Fakult\"at f\"ur Physik und Astronomie, Universit\"at
W\"urzburg, D-97074 W\"urzburg, Germany}
\author{Dietrich E. Wolf}
\affiliation{Department of Physics, Universit\"at
Duisburg-Essen, D-47048 Duisburg, Germany}
                                   

\begin{abstract}
  We show that the static friction force which must be overcome to
  render a sticking contact sliding is reduced if an external torque
  is also exerted. As a test system we study a planar disk lying on
  horizontal flat surface. We perform experiments and compare with
  analytical results to find that the coupling between static friction
  force and torque is nontrivial: It is not determined by the Coulomb
  friction laws alone, instead it depends on the microscopic details
  of friction.  Hence, we conclude that the macroscopic experiment
  presented here reveals details about the microscopic processes lying
  behind friction.
\end{abstract}

\maketitle
\parskip 2mm

\section{Introduction}
%
Although the scientific investigation of friction started several hundred years
ago with the first quantitative experiments by Leonardo da
Vinci~\cite{Dowson79},   
our knowledge about the microscopic basis for friction is surprisingly
incomplete. This applies in particular to the onset of sliding, i.e.,
the transition from static to dynamic friction. For example, it is
an open question whether the contact points give way simultaneously or
sequentially on a certain time scale \cite{Mueser03,Dedkov00}.

Recently significant progress in this direction was made by Rubinstein 
{\it et al.}, who have used fast photo arrays in order to monitor the
dynamics of contact points at the onset of sliding~\cite{Rubinstein04}. 
Pushing a plexiglass slider linearly they find that the contact points 
give way in a sequence which travels from the trailing to the leading
edge of the block. The front moves initially with half of the surface
wave speed, then accelerates, and finally splits up into a sub- and an intersonic front.
Although the dynamics of front propagation is not yet fully understood, 
these experiments show that the time scale on which the contacts give way is
very short.

In this paper we study the interplay of translation and {\it rotation}
at the onset of sliding both experimentally and theoretically. To this end
we exert simultaneously a force and a torque on a planar disk lying on
a flat surface. Since
translational and rotational static friction have the same microscopic
origin they are mutually coupled.
In particular, the critical force at the onset of sliding and spinning
turns out to depend on the torque and vice versa. We argue that this critical
line of forces and torques, where the disk starts moving, reveals
information about the microscopic dynamics, which is not as easily accessible
in experiments using linearly moving sliders.

The interplay of force and torque for a {\it sliding} disk was studied
previously in Refs.~\cite{ruina,howe,zenoprl}. It was shown that the sliding
friction of a circular disk is reduced if the contact is also spinning
with relative angular velocity $\qo$ -- a phenomenon which plays
an important role in various games such as curling or ice hockey
\cite{voyenli,sh2,nature03}.
It turns out that this reduction depends on the dimensionless ratio
$\qe =v/\qo R$,
where $R$ denotes the radius of the disk and $v$ is the
tangential relative velocity at the center of the contact area.
Based on the Coulomb friction law one obtains a sliding friction
force
\begin{equation}
|\mathbf{F}|=\mu_{\rm d} N {\qF}(\qe)
\end{equation}
and a friction torque
\begin{equation}
|\mathbf{T}|=\mu_{\rm d} N R {\qT}(\qe),
\end{equation}
where $\mu_{\rm d}$ is the dynamic friction coefficient and $N$ is
the integrated normal force acting on the contact area. Apart from the limit
of pure sliding $\qe \to \infty$, where ${\qF} \rightarrow 1$,
the functions ${\qF(\qe)}$ and ${\qT(\qe)}$ depend on the pressure
distribution across the contact area \cite{ruina,howe,sh3}. Assuming
uniform pressure
over the area of the disk these functions have been evaluated
analytically, describing the coupling of force and torque of a
circular disk in the sliding case \cite{zenoprl}.

Turning to {\it static} friction let us now consider a resting disk.
Applying simultaneously a torque and a force we are interested in the
thresholds $(\mathbf{F_{c}},\mathbf{T_{c}})$ at which the disk
starts moving. Our daily experience tells us that if we want to move a heavy
object across the floor it is easier to do so if we apply a torque to it while
pushing. But how are these quantities, force and torque, exactly related?
The aim of this paper is to determine this relation experimentally and to study
possible theoretical implications with respect to the microscopic aspects of friction.

Regarding the microscopic dynamics, the advantage of friction experiments
involving rotational degrees of freedom lies in the fact that stresses
at the contact points of the surface with the underlying support are not
evenly distributed under simultaneous action of a torque and a force.
Therefore, the question arises as to how sliding and spinning set in.
Intuitively one may think of two possible scenarios:
\begin{enumerate}
\item[(a)]
When the threshold is reached at those microcontacts where the local stress is maximal,
these contacts may break \textit{irreversibly}. After breaking the released stress is
distributed among the remaining microcontacts. As some of these
contacts cannot sustain the increasing stress anymore and break, an
avalanche-like process sets in so that eventually all contacts break and the
whole disk begins to move.
\item[(b)]
As a different scenario, the broken microcontacts may
immediately rearrange themselves to form new contacts, redistributing the
released stress over the remaining {\it and} the newly
formed contact points. This microscopic stick-slip creeping continues until
all contact points self-organize in such a way that they
sustain approximately the same stress. Therefore, by increasing
the external force or torque, all microcontacts of a perfectly rigid slider
reach the threshold of detachment simultaneously.
\end{enumerate}

We note that experiments such as those of Rubinstein {\it et al.}~\cite{Rubinstein04},
which do not involve rotational degrees of freedom,
cannot easily discriminate between the two scenarios. Although the existence
of propagating fronts in these experiments seems to favor scenario (a),
the high propagation velocity indicates that these fronts may be caused by
the inherent elasticity of the slider, leading to slightly higher stresses
of the microcontacts at the trailing edge.

In the following section we report on experimental results for a disk
subjected to an external force and torque, determining the critical line
at which spinning and sliding set in.
In Section ~\ref{TheorySection} we study simple microscopic models based on
the two scenarios described above in order to calculate the critical line
analytically. It turns out that for the avalanche scenario (a) a linear
dependence is found while in the second case (b) a nontrivial curve is obtained.
Thus the two scenarios lead to a different measurable macroscopic coupling
between static force and torque. Comparing these results with the experimental
data we can rule out scenario (a) while we find convincing agreement with scenario (b).
Finally, in Section~\ref{OnsetSection} we discuss the dynamics of a disk
shortly after the onset of sliding.
%
%
\section{Experiments}
\label{ExperimentalSection}
%
In order to determine the critical line of detachment
we performed a series of experiments where a pulling force and a
torque were applied simultaneously to a slider on a horizontal
surface.


\begin{figure}
\centerline{\epsfig{figure=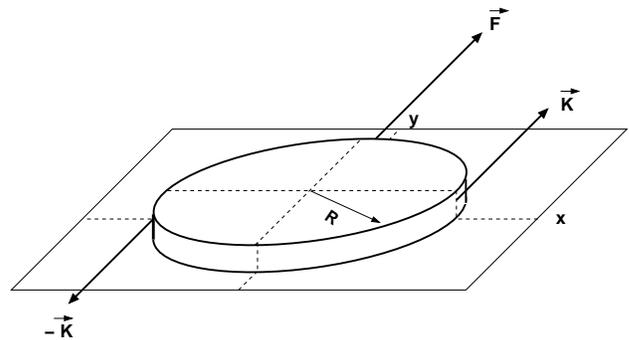,width=0.95\linewidth}}
\caption{Schematic view of the experiment: a uniform disk of
radius $R$ lies on a flat horizontal surface and is
subject to an applied
torque $\mathbf{T} = 2R\mathbf{K}$ and force $\mathbf{F}$.}
\label{fig:disk}
\end{figure}

Most of the experiments were carried out using circular disks
made of different materials with radii ranging from $R=149$ to
$R=160$ mm and masses ranging from $324$ to $2,278$ g.
All disks had mechanically polished surfaces and were provided
with small hooks along the perimeter (see Fig.~\ref{fig:disk})
from which they could be pulled.
To measure torques and forces each disk was placed on a
fixed and macroscopically flat horizontal surface covered with
carpet. Carpet--covered tracks guarantee a more uniform
pressure distribution over the contact area and have been
used successfully in other friction experiments
before~\cite{feder_feder,zenoprl}.
Force meters were then 
attached to the disks through hooks.

Once the disk was placed on the surface a torque was applied (as
indicated in Fig.~\ref{fig:disk} by the force pair $\mathbf{K}$ and
$-\mathbf{K}$). The disk was slowly pulled until it started moving.
The force meters were set to register the maximum applied pulling
force $\mathbf{F_c}$. For each fixed value of the torque a set of maximum force
readings were made. The experiments were repeated several times
under similar temperature and humidity conditions.

\begin{figure}
\centerline{\epsfig{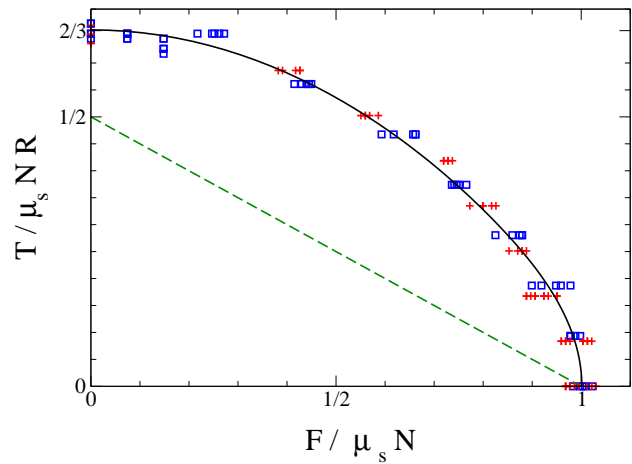}}
\caption{Measured values of torque and force for a wooden ($+$) and
plastic ($\square$) disk as a function of the dimensionless variables
$\qF=F/\mu_{s}N$ and $\qT=T/\mu_{s}NR$. The dashed and solid
lines represent the theoretical predicitions for scenarios (a) and
(b), respectively
(see text for details).}
\label{fig:experiment}
\end{figure}

In Fig.~\ref{fig:experiment} we present, for the sake of
clarity, the results of two selected experiments
using disks made of wood and plastic (disks
of different materials such as brass or steel with
different weights and sizes lead to similar results).
The curves are parameterized in terms of the dimensionless variables
\begin{equation}
\qF=\frac{|\mathbf{F}|}{\mu_sN}\,, \qquad
\qT=\frac{|\mathbf{T}|}{\mu_sNR}\,.
\end{equation}
where $N$ is the normal load, and $\mu_s$ is the static friction coefficient which is determined such that
the average over the measurements, $\langle \qF \rangle = 1$ at
the threshold from sticking to sliding without torque. As can be seen,
the experimental results are
in excellent agreement with the theoretical prediction for scenario (b),
which is shown as a solid line and will be derived in the following section.

For small $\qF$ the measurement described above is difficult to perform
since the applied torques are already close to their threshold value
without additional applied force.
Therefore, we inverted the procedure for small $\qF$,
\textit{i.e.} forces were kept fixed and torques were varied
until the critical threshold was reached.
Control experiments confirmed that both types of measurement
give compatible results within experimental error.

\section{Theory}
\label{TheorySection}

In order to determine the static thresholds of force and torque analytically
for the two scenarios described in the Introduction, let us
consider a simple model in which the microcontacts below
the threshold may be thought of as elastic springs.
This means that external forces, which are too weak to let the disk slide,
are compensated by tiny elastic deformations of microcontacts. These
deformations cause a measurable recoil (translation $\delta y$ and
rotation $\delta \varphi$), when the external forces are switched off.
The recoil was observed by placing a small mirror
on the surface of the disk and letting a laser beam reflect on it.
The beam was project onto a screen a few meters away so as make
the small displacements visible.

The local displacement of
a coarse grained surface element of the disk at a distance $r$ from
the center can be expressed by
\begin{equation}
\mathbf{u}(r,\varphi) = \delta y \,\mathbf{e}_y + r\,
\delta \varphi \, \mathbf{e}_{\varphi},
\label{eq:u}
\end{equation}
where $\delta y$ denotes the displacement of the disk and $\delta
\varphi$ is the rotation
angle with respect to the center of the disk. We assume that in a
coarse grained description the elastic restoring force per unit area is
\begin{equation}
\mathbf{f}(\mathbf{r})= -k \,\mathbf{u}(\mathbf{r})\;.
\label{neweq:f}
\end{equation}

In the case of a circular disk
the external force and torque
are given by
\begin{eqnarray}
\mathbf{F} &=& -\int_{\mathbf{r}\in A}d^2 r \; \mathbf{f}(\mathbf{r}),
\label{neweq:F}
\\
\mathbf{T} &=& -\int_{\mathbf{r}\in A}d^2 r \;
\mathbf{r}\times\mathbf{f}(\mathbf{r}),
\label{neweq:T}
\end{eqnarray}
where the integrals are performed over the area $A$ of the disk.
As long as the slider does not yet move, the local restoring forces
integrated over
the contact area compensate the external force $\mathbf{F}$ and
the external torque $\mathbf{T}$.

\begin{figure}
\centerline{\epsfig{figure=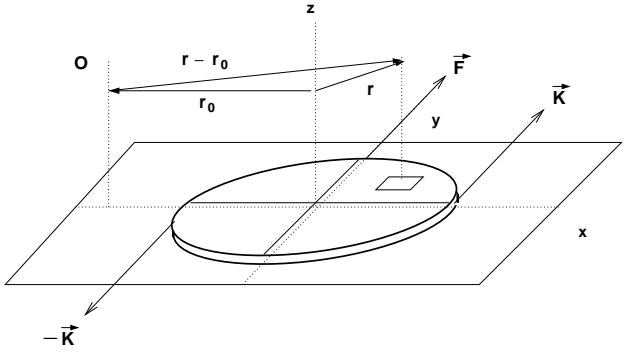,width=0.95\linewidth}}
\caption{Geometry used to solve the equations for the coupling of
forces and torques. $\mathbf{r}$ is the position of an infinitesimal
element of the disk and $\mathbf{r_0}$ the position vector of the
instantaneous center of rotation $O$.}
\label{fig:disk2}
\end{figure}

We recall that combined translation and rotation of a rigid body in a
plane can be interpreted at every moment as a pure rotation around
a particular point $\mathbf{r}_0 = -\frac{\delta y}{\delta\varphi}\mathbf{e}_x$
(see Fig. \ref{fig:disk2}), so that
\begin{equation}
\mathbf{u}(\mathbf{r}) = (\mathbf{e}_z\times(\mathbf{r}-\mathbf{r}_0))
\delta\varphi.
\label{neweq:u}
\end{equation}
For later convenience we introduce the dimensionless parameter
\begin{equation}
\gamma \equiv \frac{r_0}{R} = \frac{\delta y}{R \delta \varphi}\;.
\label{neweq:gamma}
\end{equation}

In what follows we assume that a microcontact breaks whenever the
local elastic force $|\mathbf{f}(\mathbf{r})|$ exceeds the threshold
$\mu_{\rm s} p$, where $\mu_{\rm s}$ is the static friction coefficient
and $p=N/\pi R^2$ is the lateral pressure. It is assumed that the
pressure and the friction coefficient are constant throughout the
contact area.

\paragraph{First scenario: Breaking of the weakest microcontact.}
In this case the disk starts sliding as soon as there exists
a contact point $\mathbf{r}$ which exceeds the threshold
$|\mathbf{f}(\mathbf{r})| = \mu_s p$ triggering an avalanche
in which all other points exceed the threshold, too.
Obviously the point $\mathbf{r} = R\mathbf{e}_x$ at the border
of the disk has the largest displacement so that it is the first
to reach the threshold. Therefore, the critical rotation
angle $\delta\varphi_{\rm c}$ is given by
\begin{equation}
\delta\varphi_{\rm c} = \frac{\mu_{\rm s} p}{k R (1+\gamma)}.
\end{equation}
Inserting this result into Eqs. (\ref{neweq:u}) and (\ref{neweq:f}), the force and
torque thresholds, (\ref{neweq:F}) and (\ref{neweq:T}), become
\begin{eqnarray}
\mathbf{F}_c &=& \frac{\gamma}{1+\gamma} \mu_{\rm s} N
\mathbf{e}_y,\\
\mathbf{T}_c &=& \frac{1}{2}\frac{1}{1+\gamma}\mu_{\rm s}N R
\mathbf{e}_z.
\end{eqnarray}
The normalized torque threshold $\qT = T_{\rm c}/\mu_{\rm s} N
R$ is therefore a {\em linear} function of the normalized force threshold
$\qF = F_{\rm c}/\mu_{\rm s} N$:
\begin{equation}
\qT = \frac{1}{2}(1-\qF)
\end{equation}

\paragraph{Second scenario: Collective breaking of microbonds.}
The previous result is in marked contrast to the second scenario, where we assume
that the forces per unit area, eq. (\ref{neweq:f}), relax and are redistributed
among existing and newly formed microcontacts, thereby self-organizing into a state
where virtually all surface elements reach the threshold
simultaneously. Since the direction of the displacement in a given point
is always the same, we assume that the {\em direction} of the local force
$\mathbf{u}/|u|$, does not change during this self-organization process, hence
\begin{equation}
\mathbf{F_c}=\mu_{\rm s} p
\int_{\mathbf{r}\in A}d^2 r
\frac{\mathbf{u}(\mathbf{r})}{|\mathbf{u}(\mathbf{r})|}
\end{equation}
and
\begin{equation}
\mathbf{T_c}=\mu_{\rm s} p
\int_{\mathbf{r}\in A}d^2 r\;\;
\mathbf{r}\times
\frac{\mathbf{u}(\mathbf{r})}{|\mathbf{u}(\mathbf{r})|}
\end{equation}
Together with eq.~(\ref{eq:u}) these integrals are exactly the same as in the
sliding and spinning case \cite{zenoprl}. One may write them in a more transparent
way in terms of the previously defined angle $\varphi$
\begin{equation}
\mathbf{F_c}= \mu_{\rm s} p \mathbf{e_y}
\int_{\mathbf{r}\in A}d^2 r
\;\;\frac{r\cos\varphi - r_0}{\sqrt{r^2 +r_0^2 -2rr_0\cos\varphi}}
\end{equation}
and
\begin{equation}
\mathbf{T_c} =  \mu_{\rm s} p \mathbf{e_z}
\int_{\mathbf{r}\in A}d^2 r
\;\;\frac{r^2-rr_0\cos\varphi}{\sqrt{r^2 + r_0^2 -2rr_0\cos\varphi}}.
\end{equation}
These integrals can be solved exactly and have been shown to depend
on $r_0$ only  through the dimensionless ratio $\gamma = r_0/R$.
The result for $\qF = F_{\rm c}/\mu_{\rm s}N$ and $\qT = T_{\rm c}/\mu_{\rm s} N R $ are
\begin{eqnarray}
\qF(\gamma)&=&\frac{2(1+\gamma)}{3\pi\gamma}
\left[ (1+\gamma^2)E\left(\frac{2\gamma^{1/2}}{1+\gamma}\right)\right.\nonumber\\
&&\;\;\;\;\;\;\;\;\;\;\;\;+\left. (1-\gamma)^2K\left(\frac{2\gamma^{1/2}}{1+\gamma}\right)\right]\\
\qT(\gamma)&=&
\frac{4(1+\gamma)}{9\pi}
\left[(2-\gamma^2) E\left(\frac{2\gamma^{1/2}}{1+\gamma}\right)\right. \nonumber\\
&&\;\;\;\;\;\;\;\;\;\;\;+\left. (1-\gamma)^2 K\left(\frac{2\gamma^{1/2}}{1+\gamma}\right)\right]
\end{eqnarray}

Here $K$ and $E$ are the complete
elliptic integrals of the first and the second kind,
respectively~\cite{gradshteyn}.
Although expressed in a more compact form,
these two formulas coincide exactly with those for sliding
friction \cite{zenoprl}, \cite{comment}.
By varying
$\gamma$ between $0$ and $\infty$ they provide a parameter
representation of the critical curve, which is shown in Fig.~\ref{fig:experiment}.
Obviously, the curve is in agreement with the experimental data,
which renders the scenario of collective breaking as the phyiscally
correct one.

\section{Onset of sliding}
\label{OnsetSection}

\begin{figure}
\centerline{\epsfig{figure=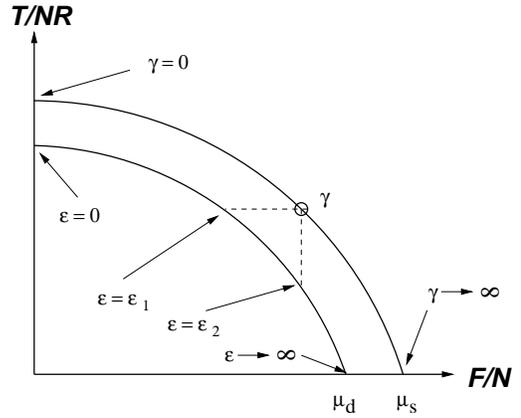,width=0.8\linewidth}}
\caption{Schematic plot of (normalized) friction force and torque.
The upper curve marks the threshold, below which a static contact is
maintained. The lower curve respresents the sliding case. See text for
details.}
\label{fig:onset}
\end{figure}

So far we studied the critical threshold from static to sliding
friction. Let us
now turn to the {\it dynamics} of the disk immediately after the onset
of sliding.
Fig. \ref{fig:onset} shows the
thresholds for the onset of sliding, $F_c/N=\mu_{\rm s}
\qF(\gamma)$ and $T_c/NR=\mu_{\rm s}\qT(\gamma)$,
which lie on a curve parametrized by the recoil parameter $\gamma=\delta
y/R\delta \varphi$. Siding friction and torque, on the other hand,
are given by $F/N=\mu_d \qF(\varepsilon)$ and $T/NR=\mu_d
\qT(\varepsilon)$ with a smaller friction coefficient $\mu_d < \mu_s$
and the motion parameter
\begin{equation}
\varepsilon=\frac{v}{\qo R}\; .
\label{eq:epsilon}
\end{equation}
If the force $F_{\rm ext}$ and torque $T_{\rm ext}$ reach the
threshold for the transition from
sticking to sliding at a certain point $\gamma$,
the body starts moving, which in general means that
it starts sliding {\em and} spinning. Hence the question arises, which
value of $\qe$ will be selected, in other words, what sliding
friction and torque will be observed immediately after the
transition from sticking to moving.

The most plausible answer is that $\qe$ will be given by the
ratio of velocity and $R$ times the angular velocity an infinitesimal
time after the motion started, i.e. (\ref{eq:epsilon}) will be replaced
by
\begin{equation}
\qe = \frac{\dot v}{\dot \qo R}\;\;.
\label{eq:selected_epsilon}
\end{equation}
The acceleration is given by the difference between static and sliding
friction
\begin{equation}
m \dot v =  F_{\rm ext} - \mu_d N {\qF}(\qe)
= N \left(\mu_{\rm s}{\qF}(\gamma) - \mu_{\rm d}{\qF}(\qe)\right),
\label{eq:v_dot}
\end{equation}
where $m$ is the mass of the slider.
Similarly the angular acceleration is given by
\begin{equation}
\Theta \dot \qo = N R \left(\mu_{\rm s}{\qT}(\gamma) - \mu_{\rm d}{\qT}(\qe)\right),
\label{eq:omega_dot}
\end{equation}
where $\Theta$ is the moment of inertia.
Inserting these equations into (\ref{eq:selected_epsilon}) one obtains
an implicit equation for the value of $\qe$ which will be
selected, if the threshold is reached at a given value of $\gamma$:
\begin{equation}
\qe = \frac{\Theta}{m R^2}\quad \frac{\mu_{\rm s}{\qF}(\gamma)- \mu_{\rm
    d}{\qF}(\qe)}{\mu_{\rm s}{\qT}(\gamma)-\mu_{\rm d}{\qT}(\qe)} .
\label{eq:implicit_epsilon}
\end{equation}

It is useful to introduce
two special values of $\qe$  for the further discussion, which
depend on the point  $\gamma$ at which
the threshold is reached. $\qe_1(\gamma) \in
[0,\infty)$ is defined by
\begin{equation}
{\qT}(\qe_1) \equiv \min\left(\frac{\mu_{\rm s}}{\mu_{\rm
      d}}{\qT}(\gamma),{\qT}(0)\right)\ .
\label{eq:epsilon_1}
\end{equation}
Similarly, $\qe_2(\gamma) \in [0,\infty)$ is defined by
\begin{equation}
{\qF}(\qe_2) \equiv \min\left(\frac{\mu_{\rm s}}{\mu_{\rm
      d}}{\qF}(\gamma),1\right)\ .
\label{eq:epsilon_2}
\end{equation}
As $\frac{\Theta}{m R^2}\geq 0$ and $\qe \geq 0$,
Eq.(\ref{eq:implicit_epsilon}) implies that $\qe(\gamma)$ is selected
from the interval
\begin{equation}
\qe_1(\gamma) \leq
\qe(\gamma) \leq
\qe_2(\gamma).
\label{eq:epsilon_interval}
\end{equation}
%
%
\section{Conclusions}
%
In this paper we studied both experimentally and theoretically the
coupling between static friction and torque for various disks in
dry contact with a track. Our results indicate that before the onset
of sliding  broken
microcontacts between slider and track rearrange themselves to form
new contacts, releasing the stresses over the remaining contacts and
the newly formed ones. Redistributing the stresses the system self-organizes until all contacts
sustain approximately the same stress. Therefore, as the force and torque
are increased up to the threshold,
all coarse-grained surface elements reach their  detachment
thresholds simultaneously and the slider moves.

Recent experiments by Rubinstein {\it et al.} \cite{Rubinstein04}
using photoarrays to detect the time evolution of the contact area
between a pexiglass slab and a track of the same material as the
threshold is reached indicate that, in the presence of a pushing force
only, the process of detachment is accompanied by a series of
propagating cracks with three different velocities. The one which
propagates most slowly is the dominant mechanism for
detachment. These experiments indicate that an avalanche-like
detachment scenario takes place at the transition from static to
sliding friction, in contradistinction to the previous discussion.

We propose that these results can be reconciled with our findings by
considering elastic deformations of the slider. In our experimental
setup the disk could be regarded as macroscopically rigid, whereas
the pexiglass slab used in \cite{Rubinstein04} may show local stress
building up at the trailing edge when being pushed. This could be
checked experimentally by pulling at the leading edge instead of
pushing at the trailing one. It would be interesting to see, whether
the detachment fronts then move in the opposite direction.

We believe that a system under the simultaneous action of a force and
a torque represents a favorable experimental setup, since each
microcontact is subject to a different displacement, which is not the
case when only a force is applied. Therefore it would be
interesting to investigate the problem described in this paper
with the technique of Rubinstein and coworkers. In particular,
how would the propagation of cracks appear in a circular geometry?

The disk geometry we use might seem rather special. However, the
concepts presented in this paper can be generalized straightforwardly
to other contact geometries as well. An example is given in
\cite{tripod}, where a tripod instead of a disk is considered.

The authors would like to thank E.Magyari and H. Thomas for bringing
their attention to the more compact version for the formulae of $\qF$
and $\qT$. This project was partially supported by project SFB 445
Nano--Particles from the Gas Phase of the DFG.



\begin{thebibliography}{99}
\bibitem{comment} Note that there is a misprint in the second formula
for $\qT$ in \cite{zenoprl}, namely the parameter $\varepsilon$ should
be in the numerator instead of the denominator.
\end{thebibliography}
 \end{document}